\documentclass[12pt]{article}
\usepackage{amssymb,latexsym,graphicx,a4,epsfig,psfrag,here}

\newcommand{\bda}{\begin{\displaymath}\begin{array}{rl}}
\newcommand{\eda}{\end{array}\end{displaymath}}
\newcommand{\be}{\begin{equation}}
\newcommand{\ee}{\end{equation}}
\newcommand{\bdm}{\begin{displaymath}}
\newcommand{\edm}{\end{displaymath}}
\newcommand{\bea}{\begin{eqnarray}}
\newcommand{\eea}{\end{eqnarray}}
\newcommand{\no}{\nonumber \\}

\def\Mpi{M_{\pi}^2}
\def\mpi{M_{\pi}}
\def\Mpin{M_{\pi^0}^2}
\def\Mpic{M_{\pi^{\pm}}^2}
\def\mpic{M_{\pi^{\pm}}}
\def\mupi{\mu_{\pi}}
\def\mupin{\mu_{\pi^0}}
\def\mupic{\mu_{\pi^{\pm}}}
\def\mupitild{\tilde{\mu}_{\pi}}

\def\MK{M_{K}^2}
\def\mK{M_{K}}
\def\MKn{M_{K^0}^2}
\def\MKc{M_{K^{\pm}}^2}
\def\mKc{M_{K^{\pm}}}
\def\muK{\mu_{K}}
\def\muKn{\mu_{{K}^0}}
\def\muKc{\mu_{{K}^{\pm}}}
\def\muKtild{\tilde{\mu}_{{K}}}

\def\Meta{M_{\eta}^2}
\def\meta{M_{\eta}}
\def\mueta{\mu_{\eta}}

\def\Fpi{F_{\pi}^2}
\def\fpi{F_{\pi}}

\def\FK{F_K^2}
\def\fK{F_K}

\def\dpi{\Delta_{\pi}}
\def\dK{\Delta_K}

\begin{document}

\thispagestyle{empty}
\begin{flushright}
CPT-2001/P.4244 \\
\today
\end{flushright}

\vspace{2cm}

\begin{center}{\Large {\bf {Isospin Breaking in Low-Energy Charged Pion and Kaon Elastic Scattering}}}

\vspace{0.5cm}

A.~Nehme\footnote{nehme@cpt.univ-mrs.fr}

\vspace{0.3cm}
\begin{center}
\textit{Centre de Physique Th\'eorique} \\
\textit{CNRS-Luminy, case 907} \\
\textit{F-13288 Marseille Cedex 09, France}.
\end{center}

\vspace{2cm}


\begin{abstract}
We use chiral perturbation theory to evaluate the scattering amplitude for the process $\pi^+K^-\rightarrow\pi^+K^-$ 
at leading and next-to-leading orders in the chiral counting and in the presence of isospin breaking effects. We also 
discuss the influence of the latter on the combination of the $S$-wave $\pi K$ scattering lengths which is 
relevant for the $2S-2P$ energy levels shift of $K\pi$ atoms.
\end{abstract}


\footnotesize{\begin{tabular}{ll}
{\bf{Keywords:}}$\!\!\!\!$& Electromagnetic corrections, 
Scattering lengths, Pion kaon scattering,\\& Chiral perturbation theory
\end{tabular}}

\end{center}
\newpage
\renewcommand{\baselinestretch}{0.6}

\tableofcontents

\renewcommand{\baselinestretch}{1}


\setcounter{equation}{0}
\renewcommand{\theequation}{\thesection.\arabic{equation}}
\section{Introduction}
\label{sec:Introduction}

The study of hadronic atoms has become a very active field. Many experiments are devoted to measure the characteristics of such atoms 
with high precision~\cite{Adeva:1994xz, Sigg:qe, Breunlich:af, Iwasaki:1997wf, Adeva:2000vb}. These experimental results carry a theoretical interest since they permit 
a direct access to hadronic scattering 
lengths providing, in this way, valuable informations concerning the fundamental properties of QCD at low energy. For instance, the 
presently running DIRAC experiment aims at measuring the pionium lifetime $\tau$ with $10\%$ accuracy~\cite{Adeva:1994xz}. This would allow 
one to determine the 
difference $a_0^0-a_0^2$ with $5\%$ precision by means of the Deser-type~\cite{Deser:1954vq} relation~\cite{Bilenkii:zd} 
\be \label{eq:lifetime}
\tau^{-1}\propto\,\left (a_0^0-a_0^2\right )^2\,, 
\ee
where $a_l^I$ denotes the $l$-wave $\pi\pi$ scattering length in the channel with total isospin $I$. On the other hand, chiral perturbation 
theory (ChPT)~\cite{Leutwyler:1993iq, Ecker:1995gg, Pich:1998xt} predictions for the scattering lengths have reached a precision amounting to 
$2\%$~\cite{Colangelo:2000jc}. Once the final results from DIRAC are 
available, ChPT will therefore be subjected to a serious test. Before confronting the experimental determination to the ChPT prediction, 
it is desirable to get all sources of corrections to the relation (\ref{eq:lifetime}), valid in the absence of isospin 
breaking, under control. In this direction, bound states calculations were performed using different approaches, like potential scattering 
theory~\cite{Moor:ye, Gashi:1997ck}, 
$3D$-constraint field theory~\cite{Jallouli:1997ux}, Bethe-Salpeter equation~\cite{Lyubovitskij:1996mb} and non-relativistic effective 
lagrangians~\cite{Kong:1998xp, Gasser:2001un}. For a review on 
the subject and a 
comparison between the various methods we refer the reader to~\cite{Gasser:1999vf}. Within the framework of non-relativistic effective 
lagrangians, the 
correct expression of relation (\ref{eq:lifetime}) which include all isospin breaking effects at leading order (LO) and next-to-leading order 
(NLO) was found 
in~\cite{Gall:1999bn} to be
\be \label{eq:corrected lifetime}
\tau^{-1}\,=\,\frac{1}{9}\,\alpha^3\left (4\Mpic -4\Mpin -\Mpic\,\alpha^2\right )^{\frac{1}{2}}{\cal A}^2\,(1+K)\,.
\ee
In the preceding equation, $\alpha$ stands for the fine-structure constant, ${\cal A}$ and $K$ possess the following 
expansions~\cite{Gall:1999bn} in powers 
of the isospin breaking parameter $\kappa\in \left [\alpha ,\,(m_d-m_u)^2\right ]$
\bea
{\cal A} &=& -\frac{3}{32\pi}\,\textrm{Re}~A_{\textrm{\scriptsize thr.}}^{+-;00}+o(\kappa )\,, \\ 
K        &=& \frac{1}{9}\left (\frac{\Mpic}{\Mpin}-1\right )\left (a_0^0+2a_0^2\right )^2-\frac{2\alpha}{3}\,(\ln\alpha -1)
               \left (2a_0^0+a_0^2\right )+o(\kappa )\,. 
\eea
The quantity of interest, 
\be \label{eq:charged to neutral}
-\frac{3}{32\pi}\,\textrm{Re}~A_{\textrm{\scriptsize thr.}}^{+-;00}\,=\,a_0^0-a_0^2+h_1(m_d-m_u)^2+h_2\alpha\,,
\ee 
represents the real part of the $\pi^+\pi^-\rightarrow\pi^0\pi^0$ scattering amplitude at order $\kappa$, calculated at threshold within 
ChPT to any 
chiral order and from which we subtract the singular pieces behaving like $q^{-1}$ and $\ln q$ whith $q$ being the centre-of-mass 
three-momentum. While $h_1$ vanishes, the coefficient $h_2$ was calculated in~\cite{Knecht:1997jw} at ${\cal O}(e^2p^2)$ where $p$ stands 
for a typical external momentum and $e$ for the electric charge.

In the first DIRAC proposal~\cite{Adeva:1994xz}, it was also planned to measure the pionium $2S-2P$ energy levels shift $\Delta E$. The 
possibility to perform  
such a measurement was discussed in~\cite{Nemenov:vp}. A simultaneous measurement of $\tau$ and $\Delta E$ would allow to pin 
down $a_0^0$ and 
$a_0^2$ separately, since~\cite{Efimov:1985fe} 
\be \label{eq:energy levels pionium}
\Delta E\propto\,2a_0^0+a_0^2\,.
\ee
Bound states calculation of the isospin breaking corrections to (\ref{eq:energy levels pionium}) were done in~\cite{Gashi:1997ck} using 
potential scattering theory. Non-relativistic effective lagrangian 
calculations concerning 
$\Delta E$ are not available. One might however expect that they will involve the quantity 
$\textrm{Re}~A_{\textrm{\scriptsize thr.}}^{+-;+-}$ for the corresponding $\pi^+\pi^-\rightarrow\pi^+\pi^-$ process. 
Electromagnetic corrections to the scattering amplitude for the process in question have been obtained at ${\cal O}(e^2p^2)$ 
in~\cite{Knecht:01ne}. 

The proposal~\cite{Adeva:1994xz} was followed by another one extending the DIRAC project by considering the possibility of measuring the 
characteristics of $K\pi$ atoms with $20\%$ accuracy~\cite{Adeva:2000vb}. The determination of the lifetime and energy levels shift for 
$K\pi$ atoms will give access to the $S$-wave $\pi K$ scattering lengths $a_0^{1/2}$ and $a_0^{3/2}$ allowing to test ChPT in the 
three-flavor sector. Likewise, it is to be expected that the relevant quantities in the final expressions for the lifetime and the levels 
shift will involve the scattering 
amplitudes for $\pi^-K^+\rightarrow\pi^0K^0$ and $\pi^+K^-\rightarrow\pi^+K^-$ respectively. In the case of the former, isospin breaking 
corrections have already been discussed in~\cite{Nehme:2001wf, Kubis:2001ij}. The aim of the present work is to provide a similar 
treatement for the latter. 

The paper is organized as follows. In Sec.~\ref{sec:Elastic scattering}, the scattering amplitude for the process 
$\pi^+K^-\rightarrow\pi^+K^-$ will be calculated at NLO including isospin breaking effects and ignoring the emission of real soft photons. 
Being lengthy, the expression for the scattering amplitude is displayed in App.~\ref{sec:Scattering amplitude}. We pursue with 
Sec.~\ref{sec:Scattering lengths} where the analytic expressions for $a_0^{1/2}$ and $a_0^{3/2}$ at NLO are given. We then discuss the 
sensitivity of the scattering lengths to the size of the next-to-next-to leading order (NNLO) by evaluating them using various inputs for 
the low-energy constants (LEC's). The threshold expansion of the process in question is performed in Sec. \ref{sec:Isospin violation} 
where the effects of isospin breaking on the scattering lengths are evaluated. Finally, App.~\ref{sec:Loop functions} collects the 
expressions for the loop functions needed in the calculation.   



\setcounter{equation}{0}
\renewcommand{\theequation}{\thesection.\arabic{equation}}
\section{Charged pion and kaon elastic scattering}
\label{sec:Elastic scattering}

The elastic scattering process 
\be \label{eq:process}
\pi^+(p_+)\,+\,K^-(k_-)\,\rightarrow\,\pi^+(p_+')\,+\,K^-(k_-')\,,
\ee
is studied in terms of the Lorentz invariant Mandelstam variables
$$
s=(p_++k_-)^2\,, \qquad t=(p_+-p_+')^2\,, \qquad u=(p_+-k_-')^2\,,
$$
verifying the on-shell relation $s+t+u\,=\,2(\Mpic +\MKc )$. These variables are related to the centre-of-mass three momentum $q$ and 
scattering angle $\theta$ by  
\bea
s &=& \left (\sqrt{\Mpic +q^2}+\sqrt{\MKc +q^2}\,\right )^2\,, \no 
t &=& -2q^2(1-\cos\theta )\,, \no 
u &=& \left (\sqrt{\Mpic +q^2}-\sqrt{\MKc +q^2}\,\right )^2-2q^2(1+\cos\theta )\,. \label{eq:c.m.f.}
\eea
Let ${\cal M}^{+-;+-}$ and ${\cal M}^{++;++}$ denote the respective scattering amplitudes for the process (\ref{eq:process}) and its 
crossed channel reaction $\pi^+K^+\rightarrow\pi^+K^+$. Then, in the isospin limit, defined by the vanishing of both the electric charge $e$ and the 
up and down quarks masses difference ($m_u=m_d\,,e=0$), the following relations hold 
\bea 
{\cal M}^{+-;+-}(s,t,u)
 &=& \frac{2}{3}\,T^{1/2}(s,t,u)+\frac{1}{3}\,T^{3/2}(s,t,u)\,, \no 
{\cal M}^{++;++}(s,t,u)
 &=& T^{3/2}(s,t,u)\,, \label{eq:isospin decomposition} 
\eea
with $T^I\,(I=1/2, 3/2)$ being the isospin amplitudes. The processes under consideration are related by $s\leftrightarrow u$ crossing 
which constrains the $T^I$ by
\be \label{eq:crossing} 
2T^{1/2}(s,t,u)\,=\,3T^{3/2}(u,t,s)-T^{3/2}(s,t,u)\,.
\ee 
This is no more valid when isospin breaking effects, generated by $\delta =m_d-m_u$ and $\alpha=e^2/(4\pi )$ are switched on. In this case 
$s\leftrightarrow u$ crossing is expressed as 
\be \label{eq:direct crossed}
{\cal M}^{+-;+-}(s, t, u)\,=\,{\cal M}^{++;++}(u, t, s)\,. 
\ee
The effect of a non-zero value for $\delta$ can be analyzed fully by means of the strong sector chiral 
lagrangian constructed in~\cite{Gasser:1985gg}. Treating isospin violation of electromagnetic origin requires the extension of ChPT in 
order to include virtual photons. This can be done by building operators in which photons figure as explicit dynamical degrees of 
freedom. The electromagnetic sector chiral lagrangian, founded upon the chiral counting scheme ${\cal O}(e)={\cal O}(p)$, was presented 
at NLO in~\cite{Urech:1995hd} and~\cite{Neufeld:1994eg}. 

\begin{figure} 
\begin{center}
\includegraphics{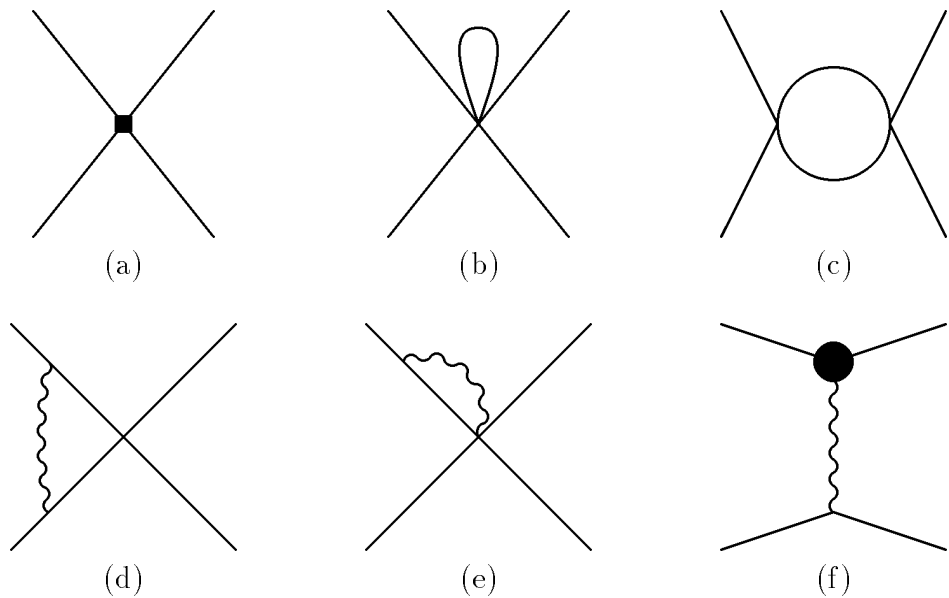}
\caption{\label{fig:2}The various types of Feynman diagrams encountered in the charged $\pi K$ scattering to 
one-loop order and ignoring ${\cal O}(e^4)$. The born-type diagram is represented by (a). Besides the contribution from the LEC's 
(the full square), it contains the 
tree contribution including the effects of bare masses and wave 
functions renormalization. Diagram (b) represents the tadpole-type part of the amplitude. The $s$-, $t$- and $u$-channel parts are 
schematized by diagram (c) and its crossed. The one-photon contribution to the amplitude follows from diagrams (d) and (e). We refer to  
diagram (f) as the form-factor-type where the full circle is made explicit in Fig.~\ref{fig:3}.}
\end{center}
\end{figure}

We shall calculate the scattering amplitude (\ref{eq:direct crossed}) at NLO including isospin breaking effects of both strong and 
electromagnetic origin. Consistency demands that all of the following chiral orders should be present; 
$p^2, e^2, \delta , p^4, e^2p^2, \delta p^2, e^4, \delta e^2, \delta^2$. From na\"{\i}ve dimensional estimation, we believe that the last 
three orders are beyond the accuracy we ask for and hence will be ignored in the following. Furthermore instead of $\delta$, our results 
will be expressed in terms of~\cite{Gasser:1985gg} 
$$
\epsilon\equiv \frac{\sqrt{3}}{4}\,\frac{m_d-m_u}{m_s-{\hat m}}=1.00\cdot 10^{-2}\,,
$$
which measures the rate of isospin violation with respect to the violation of $SU(3)$. Using 
Feynman graph techniques, the amplitude (\ref{eq:direct crossed}) can be represented at NLO by the one-particle-irreducible 
diagrams depicted in Fig.~\ref{fig:2}. 

\begin{figure} 
\begin{center}
\includegraphics{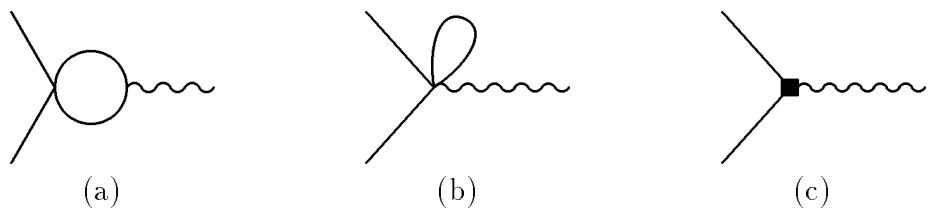}
\caption{\label{fig:3}The electromagnetic vertex function of a charged meson to one-loop order. The full square takes into account the 
contribution from the LEC's just as the tree contribution including the effect of wave function renormalization. Are shown only diagrams 
of ${\cal O}(e\,p^3)$.} 
\end{center}
\end{figure}

Although the evaluation of these diagrams is a simple exercise in quantum field theory, it involves different masses, yielding, in that 
manner, lengthy expressions. These are displayed in App.~\ref{sec:Scattering amplitude} where we keep the contributions of the various 
diagrams separated. It is useful to mention, at this stage, that these expressions are scale independent but infrared divergent. Since only 
observables are infrared safe, these infrared divergencies should cancel in the expression for the cross section where not only virtual 
photons but also real soft photons have to be taken into account. The ${\cal O}(\alpha )$ soft photon cross section corresponding to an 
arbitrary matrix element was calculated in~\cite{Denner:kt}. We cheked that by applying the general formula derived in that reference to 
the present case, 
the terms in $\ln m_{\gamma}$ so obtained cancel with the ones coming from our expressions. 



\setcounter{equation}{0}
\renewcommand{\theequation}{\thesection.\arabic{equation}}
\section{Scattering lengths}
\label{sec:Scattering lengths}

We are interested by the $S$-wave $\pi K$ scattering lengths~\cite{Weinberg:1966kf, Cronin:1967jq, Knecht:1993eq}. To this end, it is convenient to 
introduce the partial 
wave 
amplitudes $t_l^I$ defined in the $s$-channel by 
\be \label{eq:partial waves}
T^I(s,\cos\theta )\,=\,32\pi\sum_l(2l+1)P_l(\cos\theta )t_l^I(s)\,,
\ee 
where $l$ is the angular momentum of the $\pi K$ system and the $P_l$'s are the Legendre polynomials. Near
threshold the partial wave amplitudes can be parametrized in terms of the scattering lengths $a_l^I$ and slope parameters $b_l^I$. In 
the normalization (\ref{eq:partial waves}), the real part of the partial wave amplitudes reads
\be \label{threshold parameters}
\textrm{Re}~t_l^I(s)\,=\,q^{2l}\left [a_l^I+b_l^Iq^2+{\cal O}(q^4)\right ]\,.
\ee 
At NLO in ChPT no analytic expressions for the scattering lengths exist in the 
literature. This is not the case in other approaches such as heavy-kaon ChPT~\cite{Roessl:1999iu} where the expansion parameter is 
$\mpi /\mK$ in addition to $\mpi /(4\pi F_0)$ or $\mK /(4\pi F_0)$ for 
ChPT\footnote{All along this paper, $\mpi$ and $\mK$ represent respectively the pion and kaon masses in the isospin limit, 
$\meta$ the eta mass, $F_0$ the coupling of Goldston bosons to axial currents in the chiral limit.}. Obviously, at any 
order in both expansions, a matching between the two approaches is feasible. Getting analytic expressions for $a_0^{1/2}$ and 
$a_0^{3/2}$ is a straightforward matter. Using the expression of the isospin amplitude $T^{3/2}$ given in~\cite{Bernard:1991kx}, 
Eqs.~(\ref{eq:c.m.f.})-(\ref{eq:crossing}) together with (\ref{eq:partial waves}) and (\ref{threshold parameters}) lead the following 
NLO expressions for the scattering lengths
\bea
32\pi a_0^{1/2}
  &=& \frac{2\mpic\mKc}{\fpi\fK}\bigg\{\,1+\frac{4}{\fpi\fK}\,(\Mpic +\MKc )\,L_5^r \no 
  &-& \frac{1}{1152\pi^2\fpi\fK}\,\frac{1}{\MKc -\Mpic}\left [9\Mpic (11\MKc -5\Mpic )
        \ln\frac{\Mpic}{\mu^2}\right. \no 
  &+& \left.2\MKc (9\MKc -55\Mpic )\ln\frac{\MKc}{\mu^2}
        +(36\mKc^4+11\MKc\Mpic -9\mpic^4)\ln\frac{\Meta}{\mu^2}\right ]\bigg\} \no 
  &+& \frac{\Mpic\MKc}{576\pi^2\Fpi\FK}\bigg\{172+576\pi^2{\cal B}(\mKc ,\mpic )
        -192\pi^2{\cal B}(\mKc ,-\mpic ) \no 
  &+& 4608\pi^2[4(L_1^r+L_2^r)+2(L_3-2L_4^r)-L_5^r+2(2L_6^r+L_8^r)]+\frac{1}{\MKc -\Mpic}\,\times \no 
  & & \left [99\Mpic\ln\frac{\Mpic}{\mu^2}-2(67\MKc -8\Mpic )\ln\frac{\MKc}{\mu^2}
        +(24\MKc -5\Mpic )\ln\frac{\Meta}{\mu^2}\right ]\bigg\}\,, \no
32\pi a_0^{3/2}
  &=& -\frac{\mpic\mKc}{\fpi\fK}\bigg\{\,1+\frac{4}{\fpi\fK}\,(\Mpic +\MKc )\,L_5^r \no 
  &-& \frac{1}{1152\pi^2\fpi\fK}\,\frac{1}{\MKc -\Mpic}\left [9\Mpic (11\MKc -5\Mpic )
        \ln\frac{\Mpic}{\mu^2}\right. \no 
  &+& \left.2\MKc (9\MKc -55\Mpic )\ln\frac{\MKc}{\mu^2}
        +(36\mKc^4+11\MKc\Mpic -9\mpic^4)\ln\frac{\Meta}{\mu^2}\right ]\bigg\} \no 
  &+& \frac{\Mpic\MKc}{576\pi^2\Fpi\FK}\bigg\{172+384\pi^2{\cal B}(\mKc ,-\mpic ) \no 
  &+& 4608\pi^2[4(L_1^r+L_2^r)+2(L_3-2L_4^r)-L_5^r+2(2L_6^r+L_8^r)]+\frac{1}{\MKc -\Mpic}\,\times \no 
  & & \left [99\Mpic\ln\frac{\Mpic}{\mu^2}-2(67\MKc -8\Mpic )\ln\frac{\MKc}{\mu^2}
        +(24\MKc -5\Mpic )\ln\frac{\Meta}{\mu^2}\right ]\bigg\}\,. \no \label{eq:scattering lengths}
\eea
The preceding expressions were obtained by using, in the NLO contributions, the Gell$-$Mann-Okubo relation (\ref{eq:GMO}) in the 
isospin limit~\cite{Gell-Mann:1962xb}. 
$\fpi$ and $\fK$ stand for the pion and kaon decay constants respectively. Their NLO expressions in 
the isospin limit can be found in~\cite{Gasser:1985gg}. For their numerical values we will use 
$\fpi = 92.4\,\textrm{MeV}$~\cite{Holstein:ua} and 
$\fK = 1.22\fpi$~\cite{Leutwyler:1984je, Fuchs:2000hq}. The $L_i$'s 
are the LEC's weighting ${\cal O}(p^4)$ operators in the effective lagrangian of~\cite{Gasser:1985gg} and whose values are collected in 
Tab.~\ref{tab:1} with 
various experimental determinations. Finally, for compactness, we introduced the function ${\cal B}$ whose expression is displayed in 
App.~\ref{sec:Loop functions}. For historical reasons~\cite{Weinberg:1966kf}, the scattering lengths were defined in terms of the charged pion and 
kaon masses 
($\mpic = 139.570\,\textrm{MeV}, \mKc = 493.677\,\textrm{MeV}$). Furthermore, $F_0^2$ was renormalized as $\fpi\fK$ instead of $\Fpi$. 
Nevertheless, 
if one whishes to adopt the second choice for the renormalization of $F_0$, the isospin limit of Eq. (\ref{eq:renormalization}) can be 
used. 

\begin{table}[h]
\begin{center}
\vspace{0.25cm}
\begin{tabular}{cccc}
\hline
                        & set I           & set II          & set III \\
\hline 
\vspace{0.05cm}
$10^3\,L_1^r(M_{\rho})$ & $0.46\pm 0.23$  & $0.53\pm 0.25$  & $0.43\pm 0.12$ \\
\vspace{0.05cm}
$10^3\,L_2^r(M_{\rho})$ & $1.49\pm 0.23$  & $0.71\pm 0.27$  & $0.73\pm 0.12$ \\
\vspace{0.05cm}
$10^3\,L_3^r(M_{\rho})$ & $-3.18\pm 0.85$ & $-2.72\pm 1.12$ & $-2.35\pm 0.37$ \\
\vspace{0.05cm}
$10^3\,L_4^r(M_{\rho})$ & $0\pm 0.5$      & $0\pm 0.5$      & $0\pm 0.5$ \\
\vspace{0.05cm}
$10^3\,L_5^r(M_{\rho})$ & $1.46\pm 0.2$   & $0.91\pm 0.15$  & $0.97\pm 0.11$ \\
\vspace{0.05cm}
$10^3\,L_6^r(M_{\rho})$ & $0\pm 0.3$      & $0\pm 0.3$      & $0\pm 0.3$ \\
$10^3\,L_8^r(M_{\rho})$ & $1.00\pm 0.20$  & $0.62\pm 0.20$  & $0.60\pm 0.18$ \\
\hline
\end{tabular}
\end{center}
\caption{\label{tab:1}Values of the $L_i^r$'s obtained in~\cite{Amoros:2001cp} by using large $N_c$ arguments~\cite{'tHooft:1974jz} and 
fitting data 
from~\cite{Rosselet:1977pu} to ChPT predictions 
at NLO (set I) as also at NNLO (set II). Set III is the equivalent of set II with data coming from the preliminary 
analysis~\cite{Truol:2000ev} of the E865 experiment.}
\end{table}

We applied 
several checks to expressions (\ref{eq:scattering lengths}). They are scale independent, the scale dependence $\mu$ of the chiral 
logarithms is compensated by the one governing the renormalization group equations~\cite{Gasser:1985gg} of the running couplings 
$L_i^r(\mu )\equiv L_i^r$. Even so, all of our expressions will be evaluated at the scale $\mu =M_{\rho}=770\,\textrm{MeV}$. Expanding  
expressions (\ref{eq:scattering lengths}) to the fourth order in powers of $\mpic /\mKc$ we recover the combinations of scattering lengths 
obtained within the 
framework of heavy-kaon ChPT~\cite{Roessl:1999iu}. Using the same inputs as in~\cite{Bernard:1991kx} we found again their numerical 
estimates. These expressions are consistent analytically and numerically with the combinations $2a_0^{3/2}+a_0^{1/2}$ and 
$a_0^{1/2}-a_0^{3/2}$ evaluated in~\cite{Nehme:2001wf}.
 
\begin{table}[h]
\begin{center}
\begin{tabular}{ccccc}
\hline
$F_0$ renormalization & $a_0^I$     & set I               & set II              & set III \\
\hline 
                      & $a_0^{1/2}$ & $0.214\pm 0.016$    & $0.202\pm 0.018$    & $0.203\pm 0.013$ \\
$F_0^2=\Fpi$          &             &                     &                     &                  \\
\vspace{5mm}
                      & $a_0^{3/2}$ & $-0.0557\pm 0.0166$ & $-0.0660\pm 0.0185$ & $-0.0644\pm 0.0133$ \\
                      & $a_0^{1/2}$ & $0.192\pm 0.011$    & $0.177\pm 0.012$    & $0.179\pm 0.009$ \\
$F_0^2=\fpi\fK$       &             &                     &                     &                  \\
                      & $a_0^{3/2}$ & $-0.0613\pm 0.0113$ & $-0.0651\pm 0.0125$ & $-0.0644\pm 0.0090$ \\          
\hline
\end{tabular}
\end{center}
\caption{\label{tab:2}ChPT predictions for the $S$-wave $\pi K$ scattering lengths at NLO with two possible renormalization choices 
for $F_0$. Set I, set II and set III are defined in Table~\ref{tab:1}.}
\end{table}

We shall now update the numerical evaluation of~\cite{Bernard:1991kx} for the scattering lengths. All inputs have been given 
exception of the eta mass to which we assign the value $\meta = 547.30\,\textrm{MeV}$. We will evaluate $a_0^{1/2}$ and $a_0^{3/2}$ using 
each of the sets defined in Tab.~\ref{tab:1} with both renormalization choices for $F_0^2$. The interest behind this is the following. 
Since the difference between $\fpi$ and $\fK$ in the NLO pieces is of ${\cal O}(p^6)$, then any difference in the values for the 
scattering lengths due to the renormalization choice for $F_0$ could be viewed as an indication for the size of the NNLO corrections. With 
respect to the different sets of Tab.~\ref{tab:1}, set I and set II were obtained by fitting respectively 
the same experimental data to one- and two-loop ChPT predictions. It follows that the possible variation in the values for the scattering 
lengths due to the use of either the two sets measures the influence of the NNLO on the NLO which, in a 
way, reflects the rate of convergence of the chiral expansion in powers of $m_s$. While Table~\ref{tab:2} collects our numerical results 
for the scattering lengths, we will concentrate on two specific combinations, 
$2a_0^{1/2}+a_0^{3/2}$ and $a_0^{1/2}-a_0^{3/2}$. For the latter, the variations are about $\sim 9\%$ due to the renormalization choice 
for $F_0$ and $\sim 1\%\,(\sim 5\%)$ due to the choice of set II instead of set I with $F_0^2=\Fpi\,(F_0^2=\fpi\fK )$. With regard to the 
former, the renormalization choice for $F_0$ induces about 
$\sim 17\%$ of variation versus $\sim 10\%$ due to the choice of the set for both renormalization choices. Notice that up to $10\%$ 
of accuracy, the combination $a_0^{1/2}-a_0^{3/2}$ is altered neither by the renormalization choice for $F_0$ nor by the choice between 
set I and set II. 
This result has already been noticed in~\cite{Nehme:2001wf} where it was concluded that the theoretical prediction for 
$a_0^{1/2}-a_0^{3/2}$ is so clean 
that if the combination in question is measured accurately, the validity of standard three-flavor ChPT can be confirmed or not.



\setcounter{equation}{0}
\renewcommand{\theequation}{\thesection.\arabic{equation}}
\section{Isospin violation at NLO}
\label{sec:Isospin violation}

The scattering lengths are well-defined quantities in the absence of radiative corrections. When virtual photons effects have to be taken into account, 
things must be handled with care. To prevent any confusion, we will perform step by step the threshold expansion of the amplitude 
(\ref{eq:direct crossed}) with the help of Apps.~\ref{sec:Scattering amplitude} and~\ref{sec:Loop functions}. When expanded in powers of 
$q$, the real part of (\ref{eq:direct crossed}) 
can be put in an analogous form as for the $\pi\pi$ scattering case~\cite{Knecht:1997jw}  
\bea
\textrm{Re}~{\cal M}^{+-;+-}(s,t,u)
  &=& \frac{\mpic\mKc}{\fpi\fK}\,\frac{e^2}{4}\,\frac{\mu_{\pi K}}{q}+
        e^2\left (\frac{u-s}{t}\right )+
        \textrm{Re}~{\cal M}_{\textrm{\scriptsize thr.}}^{+-;+-}+{\cal O}(q)\,, \label{eqa:coulomb} \\ 
\textrm{Re}~{\cal M}^{++;++}(s,t,u)
  &=& \frac{\mpic\mKc}{\fpi\fK}\,\frac{e^2}{4}\,\frac{\mu_{\pi K}}{q}+
        e^2\left (\frac{s-u}{t}\right )+
        \textrm{Re}~{\cal M}_{\textrm{\scriptsize thr.}}^{++;++}+{\cal O}(q)\,. \label{eqb:coulomb} 
\eea
The first two terms in the right-hand side of equation (\ref{eqa:coulomb}) should be absorbed in the static characteristics of $K\pi$ 
atoms~\cite{Nemenov:1984cq, Efimov:1987jk, Belkov:1986xn} by means of a bound states treatement. The term in $q^{-1}$ is due to the Coulomb photon 
exchanged between the 
scattered particles in diagram (d) of Fig.~\ref{fig:2}. As to the term where the 
dependence on the Mandelstam variables is kept explicit, it corresponds to diagram (f) of Fig.~\ref{fig:2} at tree level and contains, 
besides the dependence on $\theta$, a singular piece behaving like $q^{-2}$. The $q$- and $\theta$-independent terms, 
$\textrm{Re}~{\cal M}_{\textrm{\scriptsize thr.}}^{+-;+-}$ and 
$\textrm{Re}~{\cal M}_{\textrm{\scriptsize thr.}}^{++;++}$, constitute the central object of the current work. Their isospin limit is 
nothing else than the threshold value of equation (\ref{eq:isospin decomposition}). Then, in order to keep a coherent notation, they will 
be denoted in the following by
$$
\textrm{Re}~{\cal M}_{\textrm{\scriptsize thr.}}^{+\mp ;+\mp}\,=\,32\pi\,a_0(+\mp ;+\mp)\,, 
$$ 
where  
\bea
a_0(+-;+-) 
  &=& \frac{1}{3}\left (2a_0^{1/2}+a_0^{3/2}\right )+\Delta a_0(+-;+-)\,, \no
a_0(++;++) 
  &=& a_0^{3/2}+\Delta a_0(++;++)\,, \label{eq:modified scattering lengths}
\eea
and the NLO expressions (\ref{eq:scattering lengths}) were used for $a_0^{1/2}$ and $a_0^{3/2}$. Let us turn towards the calculation 
of the isospin breaking quantities $\Delta a_0$. To begin with, we express the isospin limit masses $\mpi$ and $\mK$ figuring at 
NLO in terms of the charged and neutral ones by means of
\be \label{eq:dashen theorem}
\Mpi\rightarrow\Mpin\,, \qquad 2\MK\rightarrow\MKc +\MKn -\gamma (\Mpic -\Mpin )\,, 
\ee
where $\gamma$ takes into account any deviation from Dashen's theorem~\cite{Dashen:1969eg} for which $\gamma =1$. Although our results 
correspond 
to $\gamma =1$, we will use the value $\gamma =1.84$~\cite{Bijnens:1996kk} as an indicator for their sensitivity to the violation of 
Dashen theorem~\cite{Bijnens:ae}. 
Furthermore, for simplification, the modified Gell$-$Mann-Okubo relation 
\be \label{eq:GMO}
3\Meta\rightarrow \,2(\MKc +\MKn )-(2\Mpic -\Mpin )\,,
\ee
will be used everywhere except in the arguments of the chiral logarithms where the eta mass is assigned to its physical value. 
Nevertheless, the effect, on the final results, of ignoring Eq.~(\ref{eq:GMO}) will be also considered. Next, since 
by convention, the isospin limit is defined in terms of the charged pion and kaon masses, the neutral ones should be replaced according to 
$$
\Mpin\rightarrow\Mpic -\Delta_{\pi}\,, \qquad \MKn\rightarrow\MKc -\Delta_K\,. 
$$
The advantage of such a procedure is that, when expanding $a_0$ in powers of $\Delta_{\pi}$ and $\Delta_K$, the zero-th order in the 
expansion is automatically defined in terms of the charged masses and reproduces the expression for the corresponding combination of 
scattering lengths as given by (\ref{eq:modified scattering lengths}) and (\ref{eq:scattering lengths}). Although sufficient for our 
purposes, the expansion to first order in $\Delta_{\pi}$ and $\Delta_K$ is somewhat tedious especially at the level of the loop functions. 
The corresponding expansions of the latter are collected in App.~\ref{sec:Loop functions}. Having all this, the last step is achieved 
by replacing $\Delta_{\pi}$ and $\Delta_K$ in the NLO terms by their LO expressions 
$$
\Delta_{\pi}\rightarrow 2Z_0e^2F_0^2\,, \qquad \Delta_K\rightarrow 2Z_0e^2F_0^2-\frac{4\epsilon}{\sqrt{3}}\,(\MK -\Mpi )\,, 
$$
which permit the following decomposition   
\bea
32\pi\,\Delta a_0(+-;+-) 
  &\equiv & \frac{\Delta_{\pi}}{\fpi\fK}+\,\frac{\epsilon}{\sqrt{3}}\,\delta_{\epsilon}^{+-;+-}+\,2Z_0e^2F_0^2\,
              \delta_{Z_0e^2}^{+-;+-}+\,e^2\,\delta_{e^2}^{+-;+-}\,, \no 
32\pi\,\Delta a_0(++;++) 
  &\equiv & \frac{\Delta_{\pi}}{\fpi\fK}+\,\frac{\epsilon}{\sqrt{3}}\,\delta_{\epsilon}^{++;++}+\,2Z_0e^2F_0^2\,
              \delta_{Z_0e^2}^{++;++}+\,e^2\,\delta_{e^2}^{++;++}\,. \label{eq:isospin violation} 
\eea
Once more, the renormalization choice for $F_0$ was fixed to $F_0^2=\fpi\fK$ with the possibility of renormalizing as $F_0^2=\Fpi$ offered 
by the relation 
\bea
\frac{1}{\fpi\fK}
  &=& \frac{1}{\Fpi}\left\{1+\frac{4}{\Fpi}\,(\Mpic -\MKc )L_5^r\right. \no 
  &-& \left.\frac{1}{128\pi^2\Fpi}\left [5\Mpic\ln\frac{\Mpic}{\mu^2}
        -2\MKc\ln\frac{\MKc}{\mu^2}-(4\MKc -\Mpic )\ln\frac{\Meta}{\mu^2}\right ]\right. \no 
  &-& \left.\frac{\dpi}{\Fpi}\left [2L_5^r-\frac{1}{128\pi^2}\left (4+5\ln\frac{\Mpi}{\mu^2}
        -\ln\frac{\MK}{\mu^2}-\ln\frac{\Meta}{\mu^2}\right )\right ]\right. \no 
  &+& \left.\frac{\dK}{\Fpi}\left [2L_5^r-\frac{1}{128\pi^2}\left (1+\ln\frac{\MK}{\mu^2}
        +2\ln\frac{\Meta}{\mu^2}\right )\right ]\right\}\,. \label{eq:renormalization}
\eea
Notice that the consistency of the chiral power counting scheme requires that the mesons masses appearing in (\ref{eq:isospin violation}) 
be set to their isospin limit; the difference being of higher order, thus, beyond the accuracy needed for the present work. 

The isospin breaking term in (\ref{eq:isospin violation}) generated by the difference between $m_u$ and $m_d$ is given by 
\bea
\delta_{\epsilon}^{+\mp ;+\mp}
  &=& \mp\frac{\mpi\mK}{288\pi^2\Fpi\FK}\bigg\{-2304\pi^2(\MK -\Mpi )L_5^r \no 
  &+& \frac{1}{4\MK -\Mpi}[36\mK^4+563\Mpi\MK -135\mpi^4-576\pi^2\Mpi\MK{\cal B}(\mK ,\pm\mpi )] \no  
  &+& \frac{1}{\MK -\Mpi}\bigg [27\Mpi (8\MK +\Mpi )\ln\frac{\Mpi}{\mu^2} \no 
  &+& (9\mK^4-274\Mpi\MK +9\mpi^4)\ln\frac{\MK}{\mu^2}
        +(18\mK^4+4\Mpi\MK -9\mpi^4)\ln\frac{\Meta}{\mu^2}\,\bigg ]\bigg\} \no 
  &+& \frac{\Mpi\MK}{288\pi^2\Fpi\FK}\bigg\{\frac{8}{4\MK -\Mpi}\,[68\MK -19\Mpi -12\pi^2
        (32\MK -5\Mpi ){\cal B}(\mK ,\pm\mpi )] \no 
  &+& \frac{1}{\MK -\Mpi}\bigg [9(6\MK +19\Mpi )\ln\frac{\Mpi}{\mu^2} \no 
  &-& 128(\MK +\Mpi )\ln\frac{\MK}{\mu^2}+(74\MK -43\Mpi )\ln\frac{\Meta}{\mu^2}\,\bigg ]\bigg\}\,. \label{eq:epsilon piece}
\eea
The isospin violating term due to the electromagnetic difference between charged and neutral mesons masses squared was found to be 
\bea
\delta_{Z_0e^2}^{+\mp ;+\mp}
  &=& \pm\frac{\mpi\mK}{1152\pi^2\Fpi\FK}\bigg\{471\delta_{+\mp} \no 
  &-& 768\pi^2[12L_5^r-{\cal B}(\mK ,\pm\mpi )]
        -\frac{64\MK}{4\MK -\Mpi}[1-18\pi^2{\cal B}(\mK ,\pm\mpi )] \no 
  &+& \frac{1}{\MK -\Mpi}\bigg [-471\Delta_{\pi K}\delta_{-\mp}-38\MK -288\pi^2\MK\,{\cal B}(\mK ,\pm\mpi ) \no 
  &+& 27(\MK +9\Mpi )\ln\frac{\Mpi}{\mu^2}-2(161\MK +11\Mpi )\ln\frac{\MK}{\mu^2}
        +(43\MK +31\Mpi )\ln\frac{\Meta}{\mu^2}\,\bigg ]\bigg\} \no 
  &+& \frac{1}{1152\pi^2\Fpi\FK}\,\frac{\Mpi\MK}{\MK -\Mpi}\bigg [78+9216\pi^2L_5^r \no 
  &-& 864\pi^2{\cal B}(\mK ,\pm\mpi )+81\ln\frac{\Mpi}{\mu^2}+110\ln\frac{\MK}{\mu^2}
        -11\ln\frac{\Meta}{\mu^2}\,\bigg ] \no 
  &-& \frac{1}{1152\pi^2\Fpi\FK}\,\frac{\mpi^4}{\MK -\Mpi}\bigg [64 \no 
  &+& 9216\pi^2[2L_4^r+L_5^r-2(2L_6^r+L_8^r)]+63\ln\frac{\Mpi}{\mu^2}-5\ln\frac{\Meta}{\mu^2}\,\bigg ] \no 
  &-& \frac{\mK^4}{576\pi^2\Fpi\FK}\bigg\{\frac{64}{4\MK -\Mpi}\,[1+18\pi^2{\cal B}(\mK ,\pm\mpi )]-\frac{1}{\MK -\Mpi}\bigg [12 \no 
  &+& 9216\pi^2(L_4^r-2L_6^r-L_8^r)+288\pi^2{\cal B}(\mK ,\pm\mpi )
        -27\ln\frac{\MK}{\mu^2}-34\ln\frac{\Meta}{\mu^2}\,\bigg ]\bigg\}\,, \label{eq:delta piece}
\eea
with $\delta_{ij}$ representing the Kronecker symbol. Finally, virtual photons induce the following correction 
\bea
\delta_{e^2}^{+\mp ;+\mp}
  &=& \pm\frac{\mpi\mK}{144\pi^2\fpi\fK}\,\frac{1}{\MK -\Mpi}\bigg\{6[5\MK +11\Mpi -384\pi^2
        (\MK -\Mpi )L_9^r] \no 
  &-& 128\pi^2(\MK -\Mpi )(3K_1^r+3K_2^r-18K_3^r-9K_4^r+4K_5^r-5K_6^r)+\frac{3}{\MK -\Mpi}\,\times \no 
  & & \left [(11\mK^4+9\MK\Mpi +12\mpi^4)\ln\frac{\Mpi}{\mu^2}
        +(4\mK^4-39\MK\Mpi +3\mpi^4)\ln\frac{\MK}{\mu^2}\right ]\bigg\} \no 
  &-& \frac{\Mpi\MK}{16\pi^2\fpi\fK}\,\frac{1}{\MK -\Mpi}\bigg\{36-32\pi^2(2K_3^r-K_4^r-4K_{10}^r
        -4K_{11}^r) \no 
  &+& \frac{1}{\MK -\Mpi}\left [(9\MK +23\Mpi )\ln\frac{\Mpi}{\mu^2}
        -5(3\MK +4\Mpi )\ln\frac{\MK}{\mu^2}\right ]\bigg\} \no 
  &+& \frac{2\mpi^4}{3\fpi\fK}\,\frac{1}{\MK -\Mpi}\,(12K_2^r-6K_3^r+3K_4^r+K_5^r
        +7K_6^r-12K_8^r-6K_{10}^r-6K_{11}^r) \no 
  &+& \frac{\mK^4}{48\pi^2\fpi\fK}\,\frac{1}{\MK -\Mpi}\bigg [ 12 \no 
  &-& 32\pi^2(12K_2^r+K_5^r+7K_6^r-12K_8^r-18K_{10}^r-18K_{11}^r)-
        \frac{9\MK}{\MK -\Mpi}\,\ln\frac{\MK}{\mu^2}\bigg ]\,. \label{eq:pure electromagnetic piece}
\eea
It is important to stress that the virtual photons contribution, as can be seen from (\ref{eq:pure electromagnetic piece}), is safe 
from infrared divergencies. The dependence on $m_{\gamma}$ appears at higher orders of the expansions (\ref{eqa:coulomb}) and 
(\ref{eqb:coulomb}) in power of $q$. This fact was already noticed in~\cite{Knecht:1997jw}. 

For the numerical estimation of the isospin breaking effects we use $\Mpin =134.976\,\textrm{MeV}$, $\MKn =497.672\,\textrm{MeV}$ and the 
values of the $L_i^r$'s corresponding to set II of Tab.~\ref{tab:1} are 
our preferred ones. Concerning the LEC $L_9$ entering the expressions for the electromagnetic form factors of the pion and the kaon, its 
value will be taken as $L_9^r=(6.9\pm 0.7)\cdot 10^{-3}$. With respect to the LEC's in the electromagnetic sector, their central values 
correspond 
to the ones quoted in~\cite{Baur:1996ya} where large $N_c$ arguments~\cite{'tHooft:1974jz} and resonances saturation were applied to pin 
them down. Moreover, an uncertainty of $\pm 1/(16\pi^2)$, coming from na\"{\i}ve dimensional 
analysis, will be attributed to each of the $K_i^r$'s. In order to compare the order of magnitudes for the scattering lengths and the 
isospin breaking corrections to them at NLO we recall the following combinations from Tab.~\ref{tab:2} corresponding to the renormalization 
choice $F_0^2=\fpi\fK$,
\be \label{eq:comparison}
\frac{1}{3}\left (2a_0^{1/2}+a_0^{3/2}\right )\,=\,0.096\,\pm\,0.012\,, \qquad a_0^{3/2}\,=\,-0.0651\pm 0.0125\,.
\ee
Having all this, we obtain for the process  
$\pi^-K^+\rightarrow\pi^-K^+$ the following estimation for the isospin breaking effects
\bea
\begin{array}{cccc}
\displaystyle\frac{1}{32\pi} & \displaystyle\frac{\Delta_{\pi}}{\fpi\fK} & = & 1.2\;10^{-3}\,, \no \\
\displaystyle\frac{1}{32\pi} & \displaystyle\frac{\epsilon}{\sqrt{3}}\,\delta_{\epsilon}^{+-;+-} 
& = & (2.3\pm 0.1)\;10^{-4}\,, \no \\
\displaystyle\frac{1}{32\pi} & 2Z_0e^2F_0^2\,\delta_{Z_0e^2}^{+-;+-} 
& = & (1.4\pm 3.9)\;10^{-4}\,, \no \\
\displaystyle\frac{1}{32\pi} & e^2\,\delta_{e^2}^{+-;+-} 
& = & (-3.8\pm 31)\;10^{-4}\,. \\
\end{array}
\eea 
From this we deduce that 
\be \label{eq:modified scattering}
a_0(+-;+-)\,=\,0.097\,\pm\,0.013\,,
\ee
indicating that the isospin breaking effects induce a correction on the combination $2a_0^{1/2}+a_0^{3/2}$ amounting to $1\%$. As for the 
process $\pi^+K^+\rightarrow\pi^+K^+$, the results are
\bea
\begin{array}{cccc}
\displaystyle\frac{1}{32\pi} & \displaystyle\frac{\Delta_{\pi}}{\fpi\fK} 
& = & 1.2\;10^{-3}\,, \no \\
\displaystyle\frac{1}{32\pi} & \displaystyle\frac{\epsilon}{\sqrt{3}}\,\delta_{\epsilon}^{++;++} 
& = & (-1.4\pm 0.1)\;10^{-4}\,, \no \\
\displaystyle\frac{1}{32\pi} & 2Z_0e^2F_0^2\,\delta_{Z_0e^2}^{++;++} 
& = & (-3.3\pm 3.9)\;10^{-4}\,, \no \\
\displaystyle\frac{1}{32\pi} & e^2\,\delta_{e^2}^{++;++} 
& = & (12.7\pm 31)\;10^{-4}\,, \\
\end{array}
\eea 
and lead to
\be
a_0(++;++)\,=\,-0.0631\,\pm\,0.0129\,.
\ee
Although the size of the correction on $a_0^{3/2}$ is slightly bigger ($\sim 3\%$), it is the former combination, 
(\ref{eq:modified scattering}), that is expected to enter the expression for the $2S-2P$ energy levels shift for $K\pi$ atoms. Before 
concluding let us comment about the effect induced on these results by deviations from the Gell$-$Mann-Okubo relation and Dashen theorem. 
If one uses the physical eta mass instead of Eq.~(\ref{eq:GMO}), the central value in~(\ref{eq:modified scattering}) increases 
by $\sim 0.0002$. On the other hand, variations of $\gamma$ in (\ref{eq:dashen theorem}) between $1$ to $1.84$ cause deviations of order 
$\sim 10^{-5}$. Accordingly, deviations from both Gell$-$Mann-Okubo relation and Dashen theorem induce corrections which are beyond the 
accuracy we ask for.  



\setcounter{equation}{0}
\renewcommand{\theequation}{\thesection.\arabic{equation}}
\section{Conclusions}
\label{sec:Conclusion}

This work was devoted to the study at NLO of isospin breaking effects on the combination 
$2a_0^{1/2}+a_0^{3/2}$ 
of $S$-wave $\pi K$ scattering lengths which is relevant for the $2S-2P$ energy levels shift of $K\pi$ 
atoms. We first gave analytic expressions for the scattering lengths which allowed to evaluate them 
using several sets for the values of the LEC's and with both replacements $F_0^2\rightarrow \fpi\fK$ 
and $F_0^2\rightarrow \Fpi$. The particularity of the sets in question is that set I and set II for 
instance were determined by fitting the same experimental data to one- and two-loop ChPT predictions. 
We obtained for the value of the afore-mentionned combination a variation amounting to $\sim 17\%$ due 
to the choice for the replacement of $F_0^2$. As for the choice between the two sets, it induces 
$\sim 10\%$ variation. These numbers can be used as indicators for the size of the NNLO corrections as 
well as for the rate of convergence of the chiral expansion in powers of $m_s$. We pursued by performing, 
in parallel with what has been done for the $\pi\pi$ scattering case, the threshold expansion of the 
scattering amplitude for the process $\pi^+K^-\rightarrow\pi^+K^-$ in the presence of photons. After 
subtracting singular pieces, we evaluated the isospin breaking corrections to the combination in question and 
noticed two features of them: i) in spite of the fact that the NLO corrections of both strong and electromagnetic 
origin are bigger than the LO one, they cancel each others at the end. ii) Their central value represents $\sim 1\%$ 
of the NLO value for the combination $2a_0^{1/2}+a_0^{3/2}$ and the uncertainty embarrassing them is negligible with 
respect to the one coming from the LEC's in the strong sector.    



\section*{Acknowledgements}
\label{sec:thanks}

The author is grateful to M. Knecht for suggesting this study, for constant encouragement, precious comments 
and valuable advices about both the content and the presentation of the document. The author wishes 
to thank H. Sazdjian for enjoyable discussions, P. Talavera for helpful and appreciated 
comments.


\appendix      


\setcounter{equation}{0}
\renewcommand{\theequation}{\thesection.\arabic{equation}}
\section{Scattering amplitude}
\label{sec:Scattering amplitude}

The amplitude (\ref{eq:direct crossed}) will be devided into seven parts depending on the nature of the corresponding Feynman diagrams 
\bea
{\cal M}^{+-;+-}
  &=& {\cal M}^{\textrm{\scriptsize born}}+{\cal M}^{\textrm{\scriptsize tadpole}} \no
  &+& {\cal M}^{s\textrm{\scriptsize -channel}}+{\cal M}^{t\textrm{\scriptsize -channel}}+
        {\cal M}^{u\textrm{\scriptsize -channel}} \no
  &+& {\cal M}^{\textrm{\scriptsize one-photon}}+{\cal M}^{\textrm{\scriptsize form factor}}\,.
\eea
We distinguish three contributions to the born part of the amplitude (diagram (a) in figure~\ref{fig:2}) 
$$
{\cal M}^{\textrm{\scriptsize born}}\,=\,{\cal M}_{\textrm{\scriptsize tree}}^{\textrm{\scriptsize born}} 
+{\cal M}_{p^4}^{\textrm{\scriptsize born}}+{\cal M}_{e^2p^2}^{\textrm{\scriptsize born}}\,.
$$
The tree contribution, denoted by ${\cal M}_{\textrm{\scriptsize tree}}^{\textrm{\scriptsize born}}$, accounts for the four-meson vertex 
derived from the leading order (LO) lagrangian. Bare masses and mesons wave functions being renormalized, the result is the following 
\bea
{\cal M}_{\textrm{\scriptsize tree}}^{\textrm{\scriptsize born}} 
  &=& \frac{1}{2F_0^2}(\Mpic +\MKc -u)+\frac{\Delta _{\pi}}{F_0^2} \no
  &+& \frac{1}{2F_0^2}(\Mpic +\MKc -u)\bigg\{\,\frac{1}{6}\,(5\mupin +3\mueta +4\muKn +6\mupic +6\muKc ) \no
  &-& \frac{8}{F_0^2}\bigg [\,2(\Mpi +2\MK )L_4^r+(\Mpi +\MK )L_5^r\bigg ] \no
  &+& 2e^2\bigg [\,-\frac{1}{8\pi^2}-\frac{1}{16\pi^2}\bigg (\,\ln\frac{m_{\gamma}^2}{\Mpi}
        +\ln\frac{m_{\gamma}^2}{\MK}\,\bigg )-2F_0^2\mupitild -2F_0^2\muKtild \no 
  &-& \frac{4}{9}(6K_1^r+6K_2^r+5K_5^r+5K_6^r)\,\bigg ]-\frac{\epsilon}{\sqrt{3}}(\mueta -\mupi )+\frac{16}{F_0^2}\left (
        \frac{\epsilon}{\sqrt{3}}\right )(\MK -\Mpi )L_5^r\,\bigg\} \no
  &+& \frac{\MKc}{6F_0^2}\bigg\{-\frac{2}{3}\mueta +\frac{2\epsilon}{\sqrt{3}}(\mueta -\mupi )+\frac{16}{F_0^2}\left (
        \frac{\epsilon}{\sqrt{3}}\right )(\MK -\Mpi )(2L_8^r-L_5^r) \no
  &-& \frac{8}{F_0^2}\bigg [\,(\Mpi +2\MK )(2L_6^r-L_4^r)+\MK (2L_8^r-L_5^r)\,\bigg ]-e^2\bigg [\frac{1}{4\pi ^2}-6F_0^2\muKtild  \no
  &-& \frac{4}{9}(6K_1^r+6K_2^r+5K_5^r+5K_6^r-6K_7-150K_8^r-2K_9^r-20K_{10}^r-18K_{11}^r)\,\bigg ]\,\bigg\}\no
  &+& \frac{\Mpic}{6F_0^2}\bigg\{\,-\mupin +\frac{1}{3}\mueta -\frac{8}{F_0^2}\bigg [\,(\Mpi +2\MK )(2L_6^r-L_4^r)+
        \Mpi (2L_8^r-L_5^r)\,\bigg ] \no
  &-& e^2\bigg [\frac{7}{4\pi ^2}-42F_0^2\mupitild -\frac{4}{9}(6K_1^r+6K_2^r \no 
  &+& 54K_3^r-27K_4^r+5K_5^r+5K_6^r-6K_7-78K_8^r-8K_9^r-134K_{10}^r-126K_{11}^r)\,\bigg ]\,\bigg\} \no
  &+& \frac{\Delta _{\pi}}{6F_0^2}\bigg\{\,\frac{3\Mpi}{8\pi ^2F_0^2}+54\mupi +28\muK +\frac{10}{3}\mueta  \no
  &+& \frac{16}{F_0^2}\bigg [-3(\Mpi +2\MK )L_4^r-3\MK L_5^r+2(\Mpi +2\MK )L_6^r+(\Mpi +\MK )L_8^r\bigg ]\bigg\}\,. \no \label{eq:born a} 
\eea
As usually, the tadpole integrals read 
$$
\mu_P\,=\,M_P^2\,\tilde{\mu}_P\,=\,\frac{M_P^2}{32\pi^2F_0^2}\,\ln\frac{M_P^2}{\mu^2}\,,
$$
and the difference between the charged and neutral mesons masses is symbolized by 
$$
\Delta_P\,=\,M_{P^{\pm}}^2\,-\,M_{P^0}^2\,.
$$ 
Note that an infrared divergent piece appears in (\ref{eq:born a}). It comes from the charged mesons wave functions renormalization and is 
regularized by assigning a fictitious mass $m_{\gamma}$ to the photon. The counter-terms contribution 
${\cal M}_{p^4}^{\textrm{\scriptsize born}}$ issues from the NLO lagrangian of the strong sector~\cite{Gasser:1985gg} and is put in the 
following form
\bea
{\cal M}_{p^4}^{\textrm{\scriptsize born}} 
  &=& \frac{1}{F_0^4}\,\sum_{i=1}^8{\cal P}_i\,L_i^r\,,  \no 
{\cal P}_1 
  &=& 8(2\Mpic -t)(2\MKc -t)\,, \no
{\cal P}_2 
  &=& 4\bigg [\,(\Sigma _{\pi^-K^+}-s)^2+(\Sigma _{\pi^-K^+}-u)^2\,\bigg ]\,, \no
{\cal P}_3 
  &=& 2(2\Mpic -t)(2\MKc -t)+2(\Sigma _{\pi^-K^+}-s)^2\,, \no
{\cal P}_4 
  &=& -\frac{2}{3}\left [\Mpi +14\MK -\frac{24\epsilon}{\sqrt{3}}(\MK -\Mpi )\right ](2\Mpic -t)-\frac{2}{3}(13\Mpi +2\MK )(2\MKc -t) \no
  &-& \frac{4}{3}(\Mpi +2\MK )(\Sigma _{\pi^-K^+}-s)+\frac{8}{3}(\Mpi +2\MK )(\Sigma _{\pi^-K^+}-u)\,, \no
{\cal P}_5 
  &=& -\frac{2}{3}(3\Mpin +\MKc -\Delta_{\pi})(2\MKc -t)-\frac{2}{3}(\Mpin +3\MKc -3\Delta_{\pi})(2\Mpic -t) \no
  &-& \frac{8}{3}(\Mpin +\MKc -\Delta_{\pi})(\Sigma _{\pi^-K^+}-s)+\frac{4}{3}(\Mpin +\MKc -\Delta_{\pi})(\Sigma _{\pi^-K^+}-u)\,, \no
{\cal P}_6 
  &=& \frac{8}{3}\bigg [\,2M_K^4+15\Mpi\MK +M_{\pi}^4-\frac{2\epsilon}{\sqrt{3}}(2M_K^4+11\Mpi\MK -13M_{\pi}^4)\,\bigg ]\,, \no
{\cal P}_7 
  &=& 0\,, \no
{\cal P}_8 
  &=& \frac{8}{3}\bigg [\,M_K^4+6\Mpi\MK +M_{\pi}^4-\frac{4\epsilon}{\sqrt{3}}(M_K^4+2\Mpi\MK -3M_{\pi}^4)\,\bigg ]\,, \label{eq:born b}
\eea
where
$$
\Sigma_{PQ}\,=\,M_P^2\,+\,M_Q^2\,.
$$
Finally, ${\cal M}_{e^2p^2}^{\textrm{\scriptsize born}}$ represents the counter-terms contribution of ${\cal O}(e^2p^2)$. It springs from 
the NLO lagrangian in the electromagnetic sector and reads 
\bea
{\cal M}_{e^2p^2}^{\textrm{\scriptsize born}}
  &=& \frac{2e^2}{27F_0^2}(12K_1^r+12K_2^r+54K_3^r+27K_4^r+10K_5^r+10K_6^r)(\Sigma _{\pi K}-u) \no
  &-& \frac{2e^2}{27F_0^2}(6K_1^r+6K_2^r+54K_3^r+27K_4^r-4K_5^r+50K_6^r)(\Sigma _{\pi K}-s)  \no
  &-& \frac{4e^2}{27F_0^2}(3K_1^r+57K_2^r+7K_5^r+34K_6^r)(\Sigma _{\pi K}-t)  \no
  &+& \frac{4e^2}{27F_0^2}\bigg [\,3(\Mpi +\MK )K_7+3(31\Mpi +43\MK )K_8^r \no
  &+& (4\Mpi +\MK )K_9^r+(94\Mpi +91\MK )K_{10}^r+90(\Mpi +\MK )K_{11}^r\,\bigg ]\,. \label{eq:born c}
\eea
The tadpole-type part of the amplitude is furnished by diagram (b) in figure~\ref{fig:2}. Its result is shown by keeping apart individual 
contributions from each meson loop 
\bea
{\cal M}^{\textrm{\scriptsize tadpole}} 
  &=& \frac{\mupin}{36F_0^2}\bigg [\,6u-t+6\Delta _{\pi^-K^+}-24\Delta _{\pi}+\frac{2\epsilon}{\sqrt{3}}(3t+2\Mpi +4\MK )\,\bigg ] \no
  &+& \frac{\mueta}{36F_0^2}\bigg [\,6u-3t-6\Mpic +2\MKc -20\Delta _{\pi}-\frac{6\epsilon}{\sqrt{3}}(t-2\Mpi +4\MK )\,\bigg ] \no
  &+& \frac{\mupic}{18F_0^2}(12u+3t-8\Mpic -12\MKc -72\Delta _{\pi}) \no
  &+& \frac{\muKn}{9F_0^2}\bigg [\,-2t+2\MKc +\Mpic -3\Delta _{\pi}+\frac{4\epsilon}{\sqrt{3}}(\MK -\Mpi )\,\bigg ] \no
  &+& \frac{\muKc}{18F_0^2}(12u+3t-12\Mpic -8\MKc -72\Delta _{\pi})\,. \label{eq:tadpole}
\eea
Concerning the unitary corrections following from diagram (b) and its two crossed in Fig.~\ref{fig:2}, they will be 
separated according to the channel specifying each crossed diagram. Moreover, the contribution of each channel will be divided into parts 
labelled by the kind of particles propagating inside the loop. For instance, in the $s$-channel 
$$
{\cal M}^{\textrm{s\scriptsize -channel}}\,=\,{\cal M}_{\pi^-K^+}^{s\textrm{\scriptsize -channel}} 
+{\cal M}_{\pi^0K^0}^{s\textrm{\scriptsize{-channel}}}+{\cal M}_{\eta K^0}^{s\textrm{\scriptsize -channel}}\,,
$$
where
\bea
{\cal M}_{\pi^-K^+}^{s\textrm{\scriptsize -channel}}
   &=& \frac{1}{36F_0^4}\bigg\{\,6F_0^2\muKc (s-\Delta _{\pi^-K^+}+4\Delta _{\pi}) \no 
   &+& (s+\Delta _{\pi^-K^+}+6\Delta _{\pi})^2{\bar B}(\Mpic , \MKc ; s) \no
   &+& 2(s-3\Delta _{\pi^-K^+})(s+\Delta _{\pi^-K^+}+6\Delta _{\pi}){\bar B}_1(\Mpic , \MKc ; s) \no 
   &+& (s-3\Delta _{\pi^-K^+})^2{\bar B}_{21}(\Mpic , \MKc ; s) \no 
   &+& 2s\bigg [\,2\Mpic -t+4(u-t-\Delta _{\pi^-K^+})\,\bigg ]{\bar B}_{22}(\Mpic , \MKc ; s)\,\bigg\}\,, \no
{\cal M}_{\pi^0K^0}^{s\textrm{\scriptsize -channel}}
   &=& \frac{1}{8F_0^4}\bigg\{\,2F_0^2\muKn \left (1+\frac{2\epsilon}{\sqrt{3}}\right )(3s-3\Mpin -\MKn ) \no
   &+& \bigg [s-\Sigma _{\pi^0K^0}+\frac{\epsilon}{\sqrt{3}}(s-9\Mpi +7\MK )\,\bigg ]^2{\bar B}(\Mpin , \MKn ; s) \nonumber \\
   &+& 2\bigg [\,s-\Sigma _{\pi^0K^0}+\frac{\epsilon}{\sqrt{3}}(s-9\Mpi +7\MK )\,\bigg ]\times \no 
   & & \bigg [\,s-\Delta _{\pi^-K^+}+\frac{\epsilon}{\sqrt{3}}(s+3\Delta _{\pi K})\,\bigg ]{\bar B}_1(\Mpin , \MKn ; s) \no
   &+& \bigg [\,s-\Delta _{\pi^-K^+}+\frac{\epsilon}{\sqrt{3}}(s+3\Delta _{\pi K})\,\bigg ]^2{\bar B}_{21}(\Mpin , \MKn ; s) \no
   &+& 2s\bigg [\,2\MKc -t+\frac{2\epsilon}{\sqrt{3}}(t-2u+2\Mpi )\,\bigg ]{\bar B}_{22}(\Mpin , \MKn ; s)\,\bigg\}\,, \no
{\cal M}_{\eta K^0}^{s\textrm{\scriptsize -channel}}
   &=& \frac{1}{24F_0^4}\bigg\{2F_0^2\muKn \left (1-\frac{3\epsilon}{\sqrt{3}}\right )
         \bigg [3s-3\Meta -\MKn -\frac{3\epsilon}{\sqrt{3}}(3s-\MK -3\Meta )\bigg ] \no
   &+& \bigg [\,s+7\Meta -9\MKn -\frac{\epsilon}{\sqrt{3}}(3s+17\Mpi -23\MK )\,\bigg ]^2{\bar B}(\Meta , \MKn ; s) \no
   &+& 2\bigg [s+3\Delta _{\pi^-K^+}-\frac{3\epsilon}{\sqrt{3}}(s-\Delta_{\pi K})\bigg ]\times \no 
   & & \bigg [s+7\Meta -9\MKn -\frac{\epsilon}{\sqrt{3}}(3s+17\Mpi -23\MK )\bigg ]{\bar B}_1(\Meta , \MKn ; s) \no
   &+& \bigg [\,s+3\Delta _{\pi^-K^+}-\frac{3\epsilon}{\sqrt{3}}(s-\Delta_{\pi K})\,\bigg ]^2{\bar B}_{21}(\Meta , \MKn ; s) \no
   &+& 2s\bigg [4u-5t+4\Mpic -2\MKc -\frac{6\epsilon}{\sqrt{3}}(t-2u+2\Mpi )\bigg ]
         {\bar B}_{22}(\Meta , \MKn ; s)\bigg\}\,. \no \label{eq:s channel}
\eea
The loop functions figuring in (\ref{eq:s channel}) and in what follows are discussed in Appendix~\ref{sec:Loop functions}, the quantity 
$\Delta_{PQ}$ stands for 
$$
\Delta_{PQ}\,=\,M_P^2\,-\,M_Q^2\,.
$$
Similar notations hold for $t$- and $u$-channels 
\bea
{\cal M}^{t\textrm{\scriptsize -channel}}
  &=& {\cal M}_{\pi^0\pi^0}^{t\textrm{\scriptsize -channel}}+{\cal M}_{\eta\eta}^{t\textrm{\scriptsize -channel}}
        +{\cal M}_{\pi^0\eta}^{t\textrm{\scriptsize -channel}} \no
  &+& {\cal M}_{\pi^+\pi^-}^{t\textrm{\scriptsize -channel}}+{\cal M}_{K^0{\bar K}^0}^{t\textrm{\scriptsize -channel}}
         +{\cal M}_{K^+K^-}^{t\textrm{\scriptsize -channel}}\,, \no 
{\cal M}^{u\textrm{\scriptsize -channel}} 
  &=& {\cal M}_{\pi^-K^+}^{u\textrm{\scriptsize -channel}}\,,
\eea
which individual parts are given by 
\bea
{\cal M}_{\pi^0\pi^0}^{t\textrm{\scriptsize -channel}}
    &=& \frac{1}{36F_0^4}\bigg\{\,-2F_0^2\mupin \bigg [\,-5t+3\Mpin +\frac{6\epsilon}{\sqrt{3}}(-5t+3\Mpi +4\MK )\,\bigg ] \no
    &+& \frac{3}{2}(t-\Mpin )\bigg [\,3t+\frac{6\epsilon}{\sqrt{3}}(3t-4\MK )\,\bigg ]{\bar B}(\Mpin , \Mpin ; t)\,\bigg\}\,,  \no
{\cal M}_{\eta\eta}^{t\textrm{\scriptsize -channel}}
    &=& \frac{\Mpin}{72F_0^4}\bigg\{\,12F_0^2\left (1-\frac{2\epsilon}{\sqrt{3}}\right )\mueta \no
    &+& \bigg [\,9t-6\Meta -2\Mpin -\frac{2\epsilon}{\sqrt{3}}(9t+8\Mpi -20\MK )\,\bigg ]{\bar B}(\Meta , \Meta ; t)\,\bigg\}\,, \no
{\cal M}_{\pi^0\eta}^{t\textrm{\scriptsize -channel}}
    &=& -\frac{1}{12F_0^4}\bigg (\,\frac{\epsilon}{\sqrt{3}}\,\bigg )\bigg\{\,2F_0^2\mupi (5t-6\Mpi -2\Meta ) \no 
    &+& 2F_0^2\mueta (5t-4\Sigma _{\pi\eta})+(3t-4\Mpi )(3t-3\Meta -\Mpi ){\bar B}(\Mpi , \Meta ; t)\,\bigg ]\,, \no
{\cal M}_{\pi^+\pi^-}^{t\textrm{\scriptsize -channel}}
    &=& \frac{1}{18F_0^4}\bigg\{\,-\frac{3}{16\pi^2}\left (\Mpic -\frac{t}{6}\right )(s-u)+
          F_0^2\mupic \left [\,5t+24\Delta _{\pi}+3(s-u)\,\right ] \no
    &+& \bigg [\,\frac{9}{4}\,t(t+8\Delta _{\pi})-3(s-u)\bigg (\Mpic -\frac{t}{4}\bigg )\,\bigg ]{\bar B}(\Mpic , \Mpic ; t)\,, \no
{\cal M}_{K^0{\bar K}^0}^{t\textrm{\scriptsize -channel}}
    &=& \frac{1}{36F_0^4}\bigg\{\,F_0^2\muKn\left [5t-3(s-u)\right ]+\frac{3}{16\pi^2}\left (\MKn -\frac{t}{6}\right )(s-u) \no
    &+& \bigg [\,\frac{9}{4}\,t^2+3(s-u)\bigg (\MKn -\frac{t}{4}\bigg )\,\bigg ]{\bar B}(\MKn , \MKn ; t)\,\bigg\}\,, \no
{\cal M}_{K^+K^-}^{t\textrm{\scriptsize -channel}}
    &=& \frac{1}{18F_0^4}\bigg\{\,F_0^2\muKc\left [5t+3(s-u)+24\Delta _{\pi}\right ]-
          \frac{3}{16\pi^2}\left (\MKc -\frac{t}{6}\right )(s-u) \no
    &+& \bigg [\,\frac{9}{4}\,t(t+8\Delta _{\pi})-3(s-u)\bigg (\MKc -\frac{t}{4}\bigg )\,\bigg ]{\bar B}(\MKc , \MKc ; t)\,, \no 
{\cal M}_{\pi^-K^+}^{u\textrm{\scriptsize -channel}} 
  &=& \frac{1}{9F_0^4}\bigg\{-6F_0^2\muKc (\Sigma _{\pi^-K^+}-u+2\Delta _{\pi}) \no 
  &+& (2\Mpic +\MKc -u+3\Delta _{\pi})^2{\bar B}(\Mpic , \MKc ; u) \no 
  &-& 2u(2\Mpic +\MKc -u+3\Delta _{\pi}){\bar B}_1(\Mpic , \MKc ; u) \no 
  &+& u^2{\bar B}_{21}(\Mpic , \MKc ; u)+u^2{\bar B}_{22}(\Mpic , \MKc ; u)\bigg\}\,. \no
\eea
With regard to the one-photon contributions, they will be classified with respect to the topology of the exchanged photon
$$
{\cal M}^{\textrm{\scriptsize one-photon}}\,=\,{\cal M}_{\textrm{\scriptsize vertex-leg}}^{\textrm{\scriptsize one-photon}}+
{\cal M}_{s\textrm{\scriptsize -channel}}^{\textrm{\scriptsize one-photon}}+
{\cal M}_{t\textrm{\scriptsize -channel}}^{\textrm{\scriptsize one-photon}}+
{\cal M}_{u\textrm{\scriptsize -channel}}^{\textrm{\scriptsize one-photon}}\,. 
$$
We distinguish a tadpole-type contribution schematized by diagram (e) in Fig.~\ref{fig:2} and for which we found
$$
{\cal M}_{\textrm{\scriptsize vertex-leg}}^{\textrm{\scriptsize one-photon}}= \frac{2e^2}{3F_0^2}\left [(u+\Delta _{\pi K})
        \left (-6F_0^2\mupitild +\frac{1}{4\pi ^2}\right )+(u-\Delta _{\pi K})\left (-6F_0^2\muKtild +\frac{1}{4\pi ^2}\right )
        \right ]\,.
$$
We also distinguish a unitary contribution given by diagram (d) in Fig.~\ref{fig:2}. The expressions relative to the different channels 
defined by the exchanged photon topology read
\bea
{\cal M}_{s\textrm{\scriptsize -channel}}^{\textrm{\scriptsize one-photon}}
   &=&  \frac{e^2}{3F_0^2}\bigg\{\frac{1}{2}F_0^2\mupitild \,\bigg [\,3(t-u)-12(\Sigma _{\pi K}-u)+s-2\Delta _{\pi K}\,\bigg ] \no
   &+& \frac{1}{2}F_0^2\muKtild \,\bigg [\,3(t-u)-12(\Sigma _{\pi K}-u)+s+2\Mpi -6\MK \,\bigg ] \no
   &+& (s-3\Mpi -\MK ){\bar B}(\Mpi , \MK , s)+(3\Delta _{\pi K}-s){\bar B}_1(\Mpi , \MK , s) \no 
   &+& 2(\Sigma _{\pi K}-s)\bigg [\,3(u-t)+2\Sigma _{\pi K}-s\,\bigg ]G_{\pi K}^-(s)
        +8\Delta _{\pi K}(\Sigma _{\pi K}-s)G_{\pi K}^+(s) \no
   &+& 6(\Sigma _{\pi K}-u)(\Sigma _{\pi K}-s)G_{\pi K}(s)
         +6(\Sigma _{\pi K}-u)\bigg (\,\frac{1}{16\pi^2}\,\bigg )\bigg\}\,, \no
{\cal M}_{t\textrm{\scriptsize -channel}}^{\textrm{\scriptsize one-photon}}
   &=&  \frac{e^2}{6F_0^2}\bigg\{-2F_0^2\mupitild \,\bigg [\,6(\Sigma _{\pi K}-u)-\frac{3}{2}(2\Mpi -t)\,\bigg ] \no
   &-& 3(\Sigma _{\pi K}-u){\bar B}(\Mpi , \Mpi , t)-\frac{1}{16\pi^2}\left [-9(\Sigma _{\pi K}-u)-\frac{1}{2}(t-8\Mpi )\right ] \no
   &+& 6(2\Mpi -t)(\Sigma _{\pi K}-u)G_{\pi \pi}(t)-\frac{3}{2}(s-u)(4\Mpi -3t)G_{\pi \pi}^-(t)\bigg\} \no
   &+&  \frac{e^2}{6F_0^2}\bigg\{-2F_0^2\muKtild \,\bigg [\,6(\Sigma _{\pi K}-u)-\frac{3}{2}(2\MK -t)\,\bigg ] \no
   &-& 3(\Sigma _{\pi K}-u){\bar B}(\MK , \MK , t)-\frac{1}{16\pi^2}\left [-9(\Sigma _{\pi K}-u)-\frac{1}{2}(t-8\MK )\right ] \no
   &+& 6(2\MK -t)(\Sigma _{\pi K}-u)G_{KK}(t)-\frac{3}{2}(s-u)(4\MK -3t)G_{KK}^-(t)\bigg\}\,, \no
{\cal M}_{u\textrm{\scriptsize -channel}}^{\textrm{\scriptsize one-photon}}
   &=&  -\frac{e^2}{3F_0^2}\bigg\{F_0^2\mupitild \,(5u-5\MK -7\Mpi )+F_0^2\muKtild \,(5u-5\Mpi -3\MK ) \no
   &+& 2(\MK -u){\bar B}(\Mpi , \MK , u)+2u{\bar B}_1(\Mpi , \MK , u) \no
   &-& 4(\Sigma _{\pi K}-u)(2\Sigma _{\pi K}-u)G_{\pi K}^-(u)
        -4\Delta _{\pi K}(\Sigma _{\pi K}-u)G_{\pi K}^+(u) \no
   &+& 6(\Sigma _{\pi K}-u)^2G_{\pi K}(u)+6(\Sigma _{\pi K}-u)\bigg (\,\frac{1}{16\pi^2}\,\bigg )\bigg\}\,. \label{eq:elem. channels}
\eea
Finally, the result for diagram (f) in Fig.~\ref{fig:2} can be put in the following form
$$
{\cal M}^{\textrm{\scriptsize form factor}}\,=\,e^2\,\bigg (\,\frac{u-s}{t}\,\bigg )\bigg [\,1+\Gamma_{\pi}(t)+\Gamma_K(t)\,\bigg ]\,, 
$$
where $\Gamma_{\pi}$ and $\Gamma_K$ represent the electromagnetic vertex functions for the pion and kaon respectively~\cite{Gasser:1984ux}
\bea
\Gamma_{\pi}(t)
  &=& \frac{t}{F_0^2}\,\bigg [\,2L_9^r-\frac{1}{96\pi^2}\,\bigg (\,\ln\frac{\Mpi}{\mu^2}+\frac{1}{2}\ln\frac{\MK}{\mu^2}
         \,\bigg )\,\bigg ] \no
  &-& \frac{2}{3F_0^2}\bigg [(\Mpi -\frac{t}{4}){\bar B}(\Mpi , \Mpi , t)+\frac{1}{16\pi^2}\,\frac{t}{12}\bigg ] \no 
  &-& \frac{1}{3F_0^2}\bigg [(\MK -\frac{t}{4}){\bar B}(\MK , \MK , t)+\frac{1}{16\pi^2}\,\frac{t}{12}\bigg ]\,, \no
\Gamma_K(t)
  &=& \frac{t}{F_0^2}\,\bigg [\,2L_9^r-\frac{1}{96\pi^2}\,\bigg (\,\ln\frac{M_K^2}{\mu^2}+\frac{1}{2}\ln\frac{M_{\pi}^2}{\mu^2}
         \,\bigg )\,\bigg ] \no
  &-& \frac{1}{3F_0^2}\bigg [(M_{\pi}^2-\frac{t}{4}){\bar B}(\Mpi , \Mpi , t)+\frac{1}{16\pi^2}\,\frac{t}{12}\bigg ] \no
  &-& \frac{2}{3F_0^2}\bigg [(M_K^2-\frac{t}{4}){\bar B}(\MK , \MK , t)+\frac{1}{16\pi^2}\,\frac{t}{12}\bigg ]\,. \no  
\eea



\setcounter{equation}{0}
\renewcommand{\theequation}{\thesection.\arabic{equation}}
\section{Loop functions}
\label{sec:Loop functions}

The loop functions we used in our calculations follow the notations of~\cite{Passarino:1978jh} and which analytic expressions were given 
in~\cite{Nehme:2001wf}. For completeness we recall the expression for the following function
\bea 
{\cal B}(x,y)
 &=& -\frac{\sqrt{(x-y)(2x+y)}}{12\pi^2(x+y)}\times \no 
 & & \left\{\textrm{arctan}\left [\frac{\sqrt{(x-y)(2x+y)}}{2(x-y)}\right ]
       +\textrm{arctan}\left [\frac{x+2y}{2\sqrt{(x-y)(2x+y)}}\right ]\right\}\,. \label{eq:eta loop function} 
\eea
For the threshold expansion of the processes analyzed in this paper, we needed the following expressions for the two-point functions 
defined in~\cite{Nehme:2001wf} 
\bea
\textrm{Re}~{\bar B}(\Mpic , \MKc , M_{\pm}^2)
  &=& \frac{1}{16\pi^2}\left (1\mp\,\frac{\mpic\mKc}{\MKc -\Mpic}\,\ln\frac{\Mpic}{\MKc}\right )\,, \no 
\textrm{Re}~{\bar B}(\Mpin , \MKn , M_{\pm}^2)
  &=& \frac{1}{16\pi^2}\left (1\mp\,\frac{\mpic\mKc}{\MKc -\Mpic}\,\ln\frac{\Mpic}{\MKc}\right ) \no 
  &+& \frac{\Delta_{\pi}}{32\pi^2}\bigg\{\frac{2}{\MK -\Mpi}\left (2\mp\frac{\mK}{\mpi}\right ) \no 
  &+& \frac{1}{(\MK -\Mpi )^2}\,[2(2\MK +\Mpi )-(\mK\pm\mpi )^2]\ln\frac{\Mpi}{\MK}\bigg\} \no
  &-& \frac{\Delta_K}{32\pi^2}\bigg\{\frac{2}{\MK -\Mpi}\left (2\mp\frac{\mpi}{\mK}\right ) \no 
  &+& \frac{1}{(\MK -\Mpi )^2}\,[2(\MK +2\Mpi )-(\mK\pm\mpi )^2]\ln\frac{\Mpi}{\MK}\bigg\}\,, \no
\textrm{Re}~{\bar B}(\Meta , \MKn , M_{\pm}^2)
  &=& \frac{1}{16\pi^2}\,[1+16\pi^2{\cal B}(\mKc ,\pm\mpic )] \no 
  &+& \frac{1}{48\pi^2}\,\frac{1}{\MKc -\Mpic}\,(5\mKc\mp2\mpic )(2\mKc\pm\mpic )\ln\frac{\Meta}{\MKc} \no 
  &-& \frac{1}{96\pi^2}\,\frac{\Delta_{\pi}}{\MK -\Mpi}\bigg [
        12\pi^2\left (\frac{\mK\pm2\mpi}{2\mK\pm\mpi}\right ){\cal B}(\mK ,\pm\mpi ) \no 
  &+& 2\left (\frac{\mK\mp\mpi}{2\mK\mp\mpi}\right )-\left (\frac{19\MK\mp 2\mK\mpi +\Mpi}
        {\MK -\Mpi}\right )\ln\frac{\Meta}{\MK}\bigg ] \no 
  &-& \frac{1}{96\pi^2}\,\frac{\Delta_K}{\MK -\Mpi}\bigg [12\pi^2
        \left (\frac{8\mK\pm 7\mpi}{2\mK\pm\mpi}\right ){\cal B}(\mK ,\pm\mpi ) \no 
  &-& 2\left (\frac{16\mK\mp 7\mpi}{2\mK\mp\mpi}\right )+\left (\frac{37\MK\mp 2\mK\mpi -17\Mpi}
        {\MK -\Mpi}\right )\ln\frac{\Meta}{\MK}\bigg ]\,, \no 
\eea
and evaluated at
$$
M_{\pm}\,=\,\mpic\,\pm\,\mKc\,.
$$
Finally, we quote the expression of the infrared divergent three-point function whith the corresponding cut structure needed for the 
calculation of diagram (d) in Fig.~\ref{fig:2} with the various channels defined in Eq.~(\ref{eq:elem. channels})
\bea
32\pi^2\lambda_{PQ}^{\frac{1}{2}}(p^2)G_{PQ}(p^2) 
  &=& 2\,\textrm{Li}_2\left [\frac{p^2+\Delta _{PQ}+\lambda _{PQ}^{\frac{1}{2}}(p^2)}
        {p^2+\Delta _{PQ}-\lambda _{PQ}^{\frac{1}{2}}(p^2)}\right ]
        -2\,\textrm{Li}_2\left [\frac{p^2-\Delta _{PQ}-\lambda _{PQ}^{\frac{1}{2}}(p^2)}
        {p^2-\Delta _{PQ}+\lambda _{PQ}^{\frac{1}{2}}(p^2)}\right ] \no 
  &-& \left\{\ln\left [\frac{\lambda _{PQ}(p^2)}{p^2m_{\gamma }^2}\right ]
        - \frac{1}{2}\ln\left [\frac{[\Delta _{PQ}-\lambda _{PQ}^{\frac{1}{2}}(p^2)]^2-p^4}
        {[\Delta _{PQ}+\lambda _{PQ}^{\frac{1}{2}}(p^2)]^2-p^4}\right ]
        -i\pi\Theta (p^2)\right\}\times \no 
  & & \left\{\ln\left [\frac{[p^2-\lambda _{PQ}^{\frac{1}{2}}(p^2)]^2-\Delta _{PQ}^2}
        {[p^2+\lambda _{PQ}^{\frac{1}{2}}(p^2)]^2-\Delta _{PQ}^2}\right ]
        +2i\pi\Theta\left [p^2-(M_P+M_Q)^2\right ]\right\}\,. \no \label{eq:three point function}
\eea
In this expression, $\Theta$ stands for the Heaviside function, $\lambda_{PQ}$ the Lehmann function 
$$
\lambda_{PQ}(p^2)\equiv\left [p^2-(M_P+M_Q)^2\right ]\left [p^2-(M_P-M_Q)^2\right ]\,,
$$
and the dilogarithm function is defined as
$$
\textrm{Li}_2~(z)\equiv -\int_0^zdt\,\frac{\ln (1-t)}{t}\,.
$$

Notice that in the equal mass limit ($M_P=M_Q$), the function (\ref{eq:three point function}) reduces to the one given 
in~\cite{Knecht:1997jw}.
 



\end{document}